\documentclass[aps,prc,nofootinbib,twocolumn,superscriptaddress]{revtex4-1}

\usepackage{epsfig}
\usepackage{color}
\usepackage[latin1]{inputenc}
\usepackage{float,amsmath, amstext, amssymb, amsfonts}
\usepackage{graphicx}
\usepackage{ulem}

\newcommand{\be}{\begin{eqnarray}}
\newcommand{\ee}{\end{eqnarray}}

\newcommand{\benum}{\begin{enumerate}}
\newcommand{\eenum}{\end{enumerate}}

\newcommand{\dsigmap}{{\frac{\ud \sigma^\textrm{p}_\textrm{dip}}{\ud^2 \bt}}}

\newcommand{\ud}{\, \mathrm{d}}
\newcommand{\qssp}{{Q^2_\mathrm{s,p}}}

\newcommand{\rt}{{\mathbf{r}_\perp}}

\newcommand{\xt}{{\mathbf{x}_\perp}}
\newcommand{\kt}{{\mathbf{k}_\perp}}
\newcommand{\yt}{{\mathbf{y}_\perp}}
\newcommand{\bt}{{\mathbf{b}_\perp}}

\begin{document}

\title{
Event-by-event gluon multiplicity, energy density and eccentricities at RHIC and LHC
}

\author{Bj\"orn Schenke}
\affiliation{Physics Department, Brookhaven National Laboratory, Upton, NY 11973, USA}

\author{Prithwish Tribedy}
\affiliation{Variable Energy Cyclotron Centre, 1/AF Bidhan Nagar, Kolkata 700064, India}

\author{Raju Venugopalan}
\affiliation{Physics Department, Brookhaven National Laboratory, Upton, NY 11973, USA}

\begin{abstract}

The event-by-event multiplicity distribution, the energy densities and energy density weighted eccentricity moments $\epsilon_n$ (up to $n=6$) at early times in heavy-ion collisions at both RHIC ($\sqrt{s}=200\,{\rm GeV}$) and LHC ($\sqrt{s}=2.76\,{\rm TeV}$) are computed in the IP-Glasma model.  This framework combines the impact parameter dependent saturation model (IP-Sat) for nucleon parton distributions (constrained by HERA deeply inelastic scattering data) with an event-by-event classical Yang-Mills description of early-time gluon fields in heavy-ion collisions. The model produces multiplicity distributions that are convolutions of negative binomial distributions without further assumptions or parameters. The eccentricity moments are compared to the MC-KLN model; a noteworthy feature is that fluctuation dominated odd moments are consistently larger than in the MC-KLN model.

\end{abstract}

\maketitle


\section{Introduction}

Event-by-event relativistic hydrodynamic models do a good job of describing the bulk features of heavy-ion collisions at the Relativistic Heavy-Ion Collider (RHIC) and the Large Hadron Collider (LHC) \cite{Schenke:2011qd}. However, the hydrodynamic description of the collective properties of this matter is sensitive to details of the initial conditions, the spatial distributions of ``frozen gluon states"  within the incoming nucleons, the event-by-event fluctuations in the number and energy distributions of produced partons, and not least, the thermalization of the initial non-equilibrium Glasma matter formed shortly after the collision. The significant amount of data now available on moments of flow distributions and particle spectra offer the promise of being able to learn about these interesting aspects of collective Quantum Chromo-Dynamics (QCD) as well as the possibility of disentangling their properties from the hydrodynamic transport properties of a thermalized Quark-Gluon Plasma. 

In a previous letter~\cite{Schenke:2012wb}, we introduced the IP-Glasma model of early time dynamics which combines the IP-Sat (Impact Parameter Saturation Model) model~\cite{Bartels:2002cj,Kowalski:2003hm} of high energy nucleon (and nuclear) wavefunctions with the classical Yang-Mills (CYM) dynamics of the Glasma fields produced after the heavy-ion collision~\cite{Kovner:1995ja,Kovchegov:1997ke,Krasnitz:1998ns,Krasnitz:1999wc,*Krasnitz:2000gz,Lappi:2003bi}. The IP-Sat model is formally similar to the classical Color Glass Condensate (CGC) McLerran-Venugopalan (MV) model for dipole cross-sections for nucleons and nuclei~\cite{McLerran:1994ni,*McLerran:1994ka,*McLerran:1994vd} but additionally includes Bjorken $x$ and impact parameter dependence through eikonalized gluon distributions of the proton that are constrained by HERA inclusive and diffractive e+p deeply inelastic scattering (DIS) data~\cite{Kowalski:2006hc} and by the available nuclear fixed target data from EMC and E665 experiments~\cite{Kowalski:2007rw}. A key dimensionful quantity which determines the qualitative behavior of cross-sections is the saturation scale $Q_s(x,b)$ which is determined self-consistently from the dipole cross-sections. The model yields good $\chi$-squared fits\footnote{Since the work of Ref.~\cite{Kowalski:2007rw}, further data and combined analyses of the ZEUS and H1 collaboration results is available. It is important to revisit the IP-Sat fits to take this information into account. Work in this direction is in progress but is outside the scope of the present work. For nuclei, data from $d+Au$ collisions as well as future p+A and e+A collider experiments will help significantly constrain $Q_s$.} to available e+p small $x$ HERA collider data~\cite{Kowalski:2006hc} and to fixed target e+A DIS data~\cite{Kowalski:2007rw}. 

The IP-Glasma model relates the DIS constrained nuclear dipole cross-sections to the initial classical dynamics of high occupancy 
gluon ``Glasma" fields after the nuclear collision. Given an initial distribution of color charges in the high energy nuclear wavefunctions, the IP-Glasma framework computes the strong early time multiple scattering of gluon fields by event-by-event solutions of Yang-Mills equations. It is more general and accurate relative to ``$k_\perp$ factorization" models often used in the small $x$ literature, especially for the soft dynamics that characterizes hydrodynamic flow.  

The IP-Glasma model has a significant feature that is of extreme importance for event-by-event flow studies. It naturally includes the effect of several sources of quantum fluctuations that can influence hydrodynamic flow\footnote{A qualitatively similar model of the effect of initial state Glasma fluctuations (on the scale $1/Q_s$) on hydrodynamic flow moments can be found in Ref.~\cite{Gavin:2012if}.}. An important source of fluctuations, generic to all models of quantum fluctuations, are fluctuating distributions of nucleons in the nuclear wavefunctions. In addition there are fluctuations in the color charge distributions inside a nucleon. This, combined with Lorentz contraction,  results in ``lumpy'' transverse projections of color charge configurations that vary event to event. The scale of this lumpiness is given {\it on average} by the nuclear saturation scale $Q_s$ which corresponds to distance scales smaller than the nucleon size~\cite{Kowalski:2007rw}.  For each such configuration of color charges, the QCD parton model predicts dynamical event-by-event fluctuations in the multiplicities, the impact parameters and the rapidities of produced gluons~\cite{Miettinen:1978jb}. 

These dynamical fluctuations (to be distinguished from geometrical fluctuations in the nucleon positions) are captured in the CGC Glasma flux tube picture of the heavy-ion collision. In this picture, the gauge fields produced in each event after the nuclear collision are obtained by solving the CYM equations for given color charge densities in the two incoming nuclei and are therefore functionals of these densities. The n-particle gluon probability distribution can be extracted from n-particle gluon cumulants obtained by taking appropriate products of the gauge fields and averaging these over the color charge density weight functionals that are given by the CGC effective theory. An equivalent method (followed here) for generating the n-gluon multiplicity distribution is to compute single inclusive multiplicities event-by-event.

In the perturbative regime $Q_s \ll k_\perp$, it was shown that the n-particle multiplicity distribution is described by a negative binomial distribution characterized by the mean multiplicity and a parameter $k$~\cite{Gelis:2009wh}. This parameter, which is unity for a Bose-Einstein distribution and 
infinity for a Poisson distribution, is computed to be $k = \zeta \frac{N_c^2-1}{2\pi}Q_s^2 S_\perp$, where $S_\perp$ is the transverse overlap area of the collision. In Ref.~\cite{Lappi:2009xa}, it was shown from a non-perturbative computation of the double inclusive gluon distribution in the Glasma that the Glasma flux tube picture is robust. Fits based on $k_\perp$ factorization to p+p and A+A multiplicity distributions give $\zeta\sim 1/6$~\cite{Tribedy:2010ab,Tribedy:2011aa}. A value of $\zeta$ of this order also appears to be required for a reasonable estimate of the amplitude of the long range di-hadron ``ridge'' correlation in p+p collisions~\cite{Dusling:2012ig}. The event-by-event solutions of CYM equations, as will be discussed in this paper, will allow us to first obtain the important result that the NBD distribution is non-perturbatively robust,  and to determine the parameters $k$ and the mean multiplicities $\bar{n}$.

With the initial conditions generated from the IP-Sat model, the solution of the classical Yang--Mills (CYM) equations in each event 
allows one to determine the evolution of the spatial distributions of the produced matter as a function of proper time. These are characterized by the spatial eccentricity moments which can be defined as 
\begin{equation}\label{eq:eccen}
 \varepsilon_n = \frac{\sqrt{\langle r^n \cos(n\phi)\rangle^2+\langle r^n \sin(n\phi)\rangle^2}}{\langle r^n \rangle}\,,
\end{equation} 
where $\langle \cdot \rangle$ is the energy density weighted average.  These spatial eccentricity moments are in turn converted to momentum space anisotropies by hydrodynamic flow. How efficiently this is done can in detail be determined from the harmonic flow coefficients $v_n$ defined through the expansion of the azimuthal particle distribution as 
\begin{equation}
  \frac{dN}{d\phi} = \frac{N}{2\pi} \left(1 + \sum_n  (2 v_n \cos(n \phi))\right)\,.
\end{equation}
Thus if viscous hydrodynamics is applicable, and the equation of state of the hot Quark-Gluon Plasma under control, one can in principle extract information on the transport coefficients of the flow by determining {\it ab initio} the initial spatial eccentricity moments $\epsilon_n$ on the one hand and measuring the anisotropy moments $v_n$ on the other. Varying the centrality of the collision, the nuclear species available, and the energy of the collision, likely provides sufficient handles to extract the shear and bulk viscosities of hot QCD matter with some degree of confidence. 

The IP-Glasma model can be compared and contrasted with previously conceived models of the initial conditions which include various Monte Carlo based realizations of the  Glauber model (see \cite{Miller:2007ri}), and CGC inspired models such as KLN \cite{Kharzeev:2000ph,Kharzeev:2001gp}, f(actorized)KLN \cite{Drescher:2006ca,Drescher:2007ax}, and rcBK \cite{Albacete:2010ad} models. The IP-Glasma model follows all these models by sampling nucleon positions stochastically from a Woods-Saxon distribution. The nucleons' color charge distributions are constrained on the average to reproduce HERA data that provide information on the spatial and energy dependence of gluon distributions in the proton.
These color charge densitities are added at every transverse position and the total nuclear color charge density is sampled to produce the color charge distribution in a single event. Given this distribution, multi-particle production is determined event-by-event from the CYM equations.

 By way of contrast, for instance in the MC-Glauber model of Ref.~\cite{Schenke:2010rr,Schenke:2011tv}, a Gaussian distributed energy density is added for each participant nucleon. Its parameters are the same for every nucleon in every event, with the width tuned to be $0.4\,{\rm fm}$ to best describe anisotropic flow data. Unlike the IP-Glasma model, the MC-KLN models employ $k_\perp$ factorization approximations to describe gluon production on the average. Further, as noted in Ref.~\cite{Dumitru:2012yr}, they do not include NBD fluctuations on the scale 
$1/Q_s$.  Finally, in distinction to all the stated models, the IP-Glasma model does not assume instantaneous thermalization; it  includes the effects of pre-equilibrium flow through the Yang-Mills evolution until the time where the components of the stress-energy tensor  are matched to those of viscous hydrodynamics. 

Though the IP-Glasma approach is a significant improvement relative to other models with regard to including pre-equilibrium flow, we emphasize that it does not include all essential features of the pre-equilibrium stage. These include quantum fluctuation driven instabilities that change the character of the CYM  flow very significantly already by times $\tau \sim 1/Q_s$, and can generate significant amounts of flow. Work in the direction of including these initial quantum fluctuations is highly advanced already~\cite{Dusling:2011rz} and ``proof of concept" studies demonstrating their role in isotropization of $\phi^4$ theories with heavy-ion like initial conditions have been performed~\cite{Dusling:2010rm,Epelbaum:2011pc,Dusling:2012ig2}. 
It is out of the scope of this work to address these pre-equilibrium flow studies but we anticipate including them in future work. 

The paper is organized as follows. In Section 2, we discuss the IP-Glasma model which combines the IP-Sat model of nucleon gluon distributions with classical Yang-Mills evolution of a nuclear collision. Color charge distributions are sampled from a Gaussian distribution with variance $g^2\mu_A^2(\xt)$. This variance is the color charge squared per unit area constructed from the individual nucleon $g^2\mu^2$ and is proportional to $Q_{S,A}^2(\xt)$. Given the color charge distribution, the classical Yang-Mills fields for each nucleus can be determined, and a unique solution for the gauge fields immediately after each collision obtained. These then provide the initial conditions for numerical solutions of the Yang-Mills equations. The details of the numerical computation are discussed in Section 3. In Section 4, we discuss results for the following: i) the dependence of the initial multiplicity as a function of the number of participants $N_{\rm part}$ compared to data at RHIC and LHC energies, ii) the multiplicity distributions, and the comparison of these to negative binomial distributions whose  parameters have the form conjectured in the Glasma flux tube picture, iii) even and odd eccentricity moments and their comparison to those in the MC-KLN model. In the final section, we outline the continuation of this work (in a follow up paper) where event-by-event hydrodynamic equations are solved with input from the IP-Glasma model to determine the moments of flow distributions. The IP-Sat model is discussed in an appendix.

\section{The IP-Glasma model}
We will present in this section details of the IP-Glasma model first introduced in \cite{Schenke:2012wb}. The present 
discussion additionally includes time evolution with the CYM equations in 2+1 dimensions. The IP-Glasma framework uses the IP-Sat model~\cite{Bartels:2002cj,Kowalski:2003hm} to determine fluctuating configurations of color charges in two incoming highly energetic nuclei. A review of the IP-Sat model can be found in Appendix \ref{sec:ipsat}.

The first step in this framework is to determine the event-by-event nuclear color charge densities of each of the nuclei which generate the corresponding classical fields. This is done as follows. Nucleon positions in the transverse plane of both nuclei are sampled from a
Fermi distribution
\begin{equation}
  \kappa(r) = \kappa_0 \frac{1+w(r/R)^2}{1+\exp(\frac{r-R}{a})}\,,
\end{equation}
where $\kappa_0$ is the nucleon density, $R$ the nuclear radius, $a$ the skin depth and $w$ denotes deviations from a spherical shape\footnote{
For $^{197}$Au, the parameters are, $R=6.38\,{\rm fm}$, $a=0.535\,{\rm fm}$, $w=0$, and for $^{207}$Pb, 
$R=6.62\,{\rm fm}$, $a=0.546\,{\rm fm}$, $w=0$. The overall normalization $\kappa_0$ is irrelevant for sampling the positions.}.

We next determine the color charge squared per unit area of both nuclei $g^2\mu_{A(B)}^2(x,\xt)$ by summing the corresponding quantities 
$g^2\mu^2(x,\bt,\xt)$ of all individual nucleons in the nucleus. 
Here $x= \langle p_\perp\rangle/\sqrt{s}$ for zero rapidity, where $\langle p_\perp\rangle$ is the average transverse momentum
of charged hadrons in p+p collisions at a given $\sqrt{s}$ and $\xt$ is the coordinate in the plane transverse to the beam line. For
$g^2\mu^2(x,\bt,\xt)$ it determines the position of the nucleon's center, while $\bt$ is the impact parameter relative to its center.
The $g^2\mu^2(x,\bt)$ are in turn proportional to the  saturation scale $Q_{s, (p)}^2(x,\bt)$ provided by the IP-Sat dipole cross section for each nucleon as discussed in  Appendix \ref{sec:ipsat}. The exact numerical factor between $g^2\mu^2(x,\bt)$ and $Q_{s, (p)}^2(x,\bt)$ depends on the details of the calculation \cite{Lappi:2007ku}; we will discuss it further below. 

\begin{figure}[h]
 \centerline{\includegraphics[width=6.5cm]{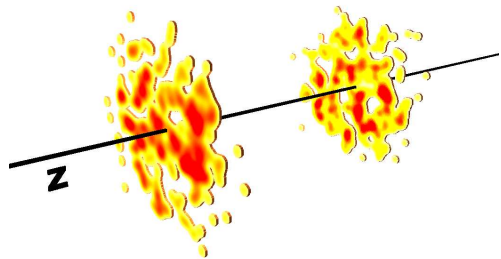}}\vspace{-0.2cm}
 \caption{The incoming color charge densities $g\mu_{A(B)}$ for two gold nuclei at $\sqrt{s}=200\,{\rm GeV}$. Increasing density from yellow to red. \label{fig:gmu}}
 \end{figure}

The distributions $g^2\mu_{A(B)}^2(x,\xt)$ are shown in Fig.\,\ref{fig:gmu}. The degree of lumpiness in this quantity is determined by the nucleon size.
Given $g^2\mu_{A(B)}^2(x,\xt)$, one can sample $\rho^a(\xt)$ in each event from the Gaussian distribution
\begin{equation}
\langle \rho_{A(B)}^a(\xt)\rho_{A(B)}^b(\yt)\rangle = g^2\mu_{A(B)}^2(x,\xt) \delta^{ab} \delta^{(2)}(\xt-\yt) \, .
\end{equation}
The Gaussian sampled color charges $\rho^a(x^\mp,\xt)$ in the IP-Sat/MV model act as local sources for small $x$ classical gluon Glasma fields. 
Here, $a$ is a color index running from 1 to $N_c^2-1$. We introduced a general dependence on $x^-$ or $x^+$, which depending on the direction the
nucleus is moving, is the longitudinal spatial light-cone coordinate. Generally, light-cone quantities are defined as $v^\pm=(v^0\pm v^3)/\sqrt{2}$, which translates to proper time and spatial rapidity as $\tau = \sqrt{2 x^+ x^-}$ and $\eta = 0.5\ln(x^+/x^-)$. This dependence allows for a finite longitudinal width of the nucleus, which we will use in the numerical calculation below.

The classical gluon fields are determined by solving the classical Yang-Mills equations 
\begin{equation}\label{eq:YM1}
  [D_{\mu},F^{\mu\nu}] = J^\nu\,,
\end{equation}
with the  color current
\begin{equation}\label{eq:current}
  J^\nu = \delta^{\nu \pm}\rho_{A (B)}(x^\mp,\xt)
\end{equation}
 generated by a nucleus A (B) moving along the $x^+$ ($x^-$) direction (the upper index is for nucleus $A$).
In (\ref{eq:current}) we have assumed that we are in a gauge where $A^\mp=0$ such that temporal Wilson lines 
along the $x^+$ ($x^-$) axis become trivial unit matrices.

The solution to Eq.\,(\ref{eq:YM1}) is most easily found in Lorentz gauge $\partial_\mu A^\mu = 0$, where the equation becomes a 
two-dimensional Poisson equation
\begin{equation}
  -\boldsymbol{\nabla}_\perp^2 A_{A(B)}^\pm = \rho_{A (B)}(x^\mp,\xt)\,,
\end{equation}
whose solution can formally be written as
\begin{equation}
  A_{A(B)}^\pm = -\rho_{A (B)}(x^\mp,\xt)/\boldsymbol{\nabla}_\perp^2\,.
\end{equation}
It will be more convenient to work in light-cone gauge $A^+ (A^-) = 0$ when computing the gluon fields after the collision.
The solution in this gauge is obtained by gauge transforming the result in Lorentz gauge using the path-ordered exponential
\begin{equation}\label{eq:wilson}
  V_{A (B)} (\xt) = P \exp\left({-ig\int dx^{-} \frac{\rho^{A (B)}(x^-,\xt)}{\boldsymbol{\nabla}_T^2+m^2} }\right)\,,
\end{equation}
giving the pure gauge fields ~\cite{McLerran:1994ni,*McLerran:1994ka,*McLerran:1994vd,JalilianMarian:1996xn,Kovchegov:1996ty}   
\begin{align}\label{eq:sol}
  A^i_{A (B)}(\xt) &= \theta(x^-(x^+))\frac{i}{g}V_{A (B)}(\xt)\partial_i V^\dag_{A (B)}(\xt)\,,\\
  A^- (A^+) &= 0\,.\label{eq:sol2}
\end{align}
The infrared regulator $m$ in Eq.\,(\ref{eq:wilson}) is of order $\Lambda_{\rm QCD}$ and incorporates color confinement at the nucleon level.
\footnote{Other prescriptions which do not explicitly introduce a mass~\cite{Krasnitz:2002mn} are feasible but they all inevitably involve introducing a nucleon size scale. This is because there is a Coulomb problem in QCD which is cured only by confinement. The presumption here is that physics at high energies is dominated by momenta $\sim Q_s$ 
and is insensitive to infrared physics at the scale $m$. From a practical point of view, we observe that our results are weakly sensitive to small variations in the scale $m$.}
Physically, the solution (\ref{eq:sol},\ref{eq:sol2}) is a gauge transform of the vacuum on one side of the light-cone and another gauge transform
of the vacuum on the other side. We have chosen one of them to be zero as an overall gauge choice. The discontinuity in the fields on the light-cone
corresponds to the localized valence charge source \cite{Kovner:1995ja}.

The initial condition for a heavy-ion collision at time $\tau=0$ is given by the solution of the CYM equations in Fock--Schwinger gauge 
$A^\tau=(x^+ A^- + x^- A^+)/\tau=0$, which is a natural choice because it interpolates between the light-cone gauge conditions of the incoming nuclei. It is also necessary
for the Hamiltonian formulation that we adopt (gauge links in the temporal ($\tau$) direction become unit matrices in this gauge). It has a simple expression in terms of the gauge fields of the colliding nuclei
\footnote{The metric in the $(\tau,\xt,\eta)$ coordinate system is $g_{\mu\nu} = {\rm diag}(1,-1,-1,-\tau^2)$ so that $A_\eta=-\tau^2 A^\eta$. The $\pm$ components of the gauge field are related by $A^\pm = \pm x^\pm A^\eta$.}\cite{Kovner:1995ja,Kovner:1995ts}:
\begin{align}
  A^i &= A^i_{(A)} + A^i_{(B)}\,,\label{eq:init1}\\
  A^\eta &= \frac{ig}{2}\left[A^i_{(A)},A^i_{(B)}\right]\,,\label{eq:init2}\\
  \partial_\tau A^i &= 0\,,\\
  \partial_\tau A^\eta &= 0
\end{align}

In the limit $\tau\rightarrow 0$, $A^\eta=-E_\eta/2$, with $E_\eta$ the longitudinal component of the electric field. At $\tau=0$, 
the only non-zero components of the field strength tensor are the longitudinal magnetic and electric fields, which can be computed non-perturbatively. 
They determine the energy density of the Glasma at $\tau=0$ 
at each transverse position in a single event~\cite{Krasnitz:1999wc,*Krasnitz:2000gz,Lappi:2003bi}.

The Glasma fields are then evolved in time $\tau$ according to Eq.\,(\ref{eq:YM1}). Over a time scale $\sim 1/Q_s$ the fields are strong and 
the system is strongly interacting. Due to the expansion of the system, the fields become weak after this time scale and the system begins to stream freely.
Incorporation of quantum fluctuations in a 3+1 dimensional CYM simulation will however lead to instabilities, which will modify this behavior and potentially keep the system strongly interacting for a more extended period of time~\cite{Romatschke:2005pm,Romatschke:2006nk}. As noted previously, these instabilities could isotropize the system, naturally leading to a transition to viscous hydrodynamic behavior.
The detailed study of instabilities and the origin of isotropization is a complex task and beyond the scope of this work. 
For recent progress in this direction see \cite{Dusling:2011rz,Blaizot:2011xf,Kurkela:2011ub,Berges:2012iw}. We emphasize that key aspects of this work, the event-by-event determination of color charge distributions and solutions of Yang--Mills equations will be essential ingredients in these generalized frameworks as well. In particular, in the framework of Ref.~\cite{Dusling:2011rz}, the additional ingredient is repeated solution of the CYM equations with slightly different seeds drawn from an initial spectrum of fluctuations. 

\section{Numerical computation}\label{sec:numcalc}

We will now discuss the numerical implementation of the continuum discussion in the previous section. Because the classical gauge field configurations are boost invariant, our computations are carried out on 2+1-dimensional lattices. 
From the nuclear color charge density squared, determined as described in the previous section, 
we can sample independent color charges $\rho^a(\xt)$ (suppressing $x$ from now on) according to
\begin{equation}
  \langle \rho_k^a(\xt)\rho_l^b(\yt)\rangle = \delta^{ab}\delta^{kl}\delta^2(\xt-\yt)\frac{g^2\mu_A^2(\xt)}{N_y}\,,
\end{equation}
where the indices $k,l=1,2,\dots,N_y$ represent a discretized $x^-$ coordinate~\cite{Lappi:2007ku}. 
We typically use $N_y=10$ and $m=\Lambda_{\rm QCD} \approx 0.2\,{\rm GeV}$ and will use a value 
of $Q_s \simeq 0.75\, g^2\mu$, which is somewhat larger than the value determined in \cite{Lappi:2007ku} for the adjoint saturation scale. 
There is some uncertainty in this relation and we have chosen $0.75$ for comparison to experimental data. Changing this value primarily changes
the overall normalization of the multiplicity and energy density.

For large nuclei, the use of local Gaussian color charge distributions is a valid approximation~\cite{McLerran:1994ni,*McLerran:1994ka,*McLerran:1994vd,Kovchegov:1996ty,Jeon:2004rk}. Modifications to Gaussian distributions, relevant for smaller nuclei, have recently been explored in \cite{Dumitru:2011zz}.

On the lattice we work with link variables $U^{i}_{A(B),j}$ instead of explicit gauge fields.
To determine $U^{i}_{A(B),j}$ we first need to compute the path ordered Wilson line (\ref{eq:wilson}) in its discretized version 
\begin{equation}
  V_{A (B)}(\xt) = \prod_{k=1}^{N_y}\exp\left(-ig\frac{\rho_k^{A (B)}(\xt)}{\boldsymbol{\nabla}_T^2+m^2}\right)\,.
\end{equation}

\begin{figure}[h]
 \centerline{\includegraphics[width=6.5cm]{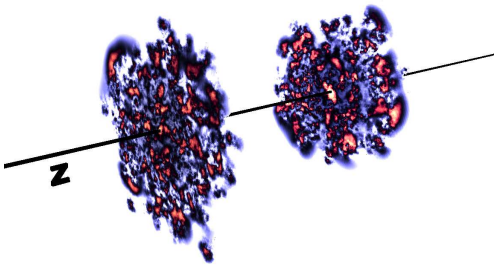}}\vspace{-0.2cm}
 \caption{The correlator $1/N_c {\rm Re} [{\rm Tr}(V^\dag_{A(B)}(0,0) V_{A(B)}(x,y))]$ showing the degree of correlations in the gluon fields
for lead ions at $\sqrt{s}=2760\,{\rm GeV}$. \label{fig:VdagVLHC}}
 \end{figure}

This quantity already allows us to demonstrate the degree of correlation and fluctuations in the gluon fields of the incoming nuclei.
Fig.\,\ref{fig:VdagVLHC} shows the correlator $ {\rm Re} [{\rm Tr}(V^\dag_{A(B)}(0,0) V_{A(B)}(x,y))]/N_c$ for $\sqrt{s}=2760\,{\rm GeV}$.
The characteristic correlation length is $1/Q_s$,
leading to a finer granularity \footnote{This energy dependence of the granularity is opposite to the one modeled in \cite{Heinz:2011mh}.}
than the nucleon size scale (cf. Fig.\,\ref{fig:gmu}).
See also \cite{Dumitru:2011vk} where the evolution of this structure with $x$ was computed by solving the JIMWLK renormalization group equations.

To each lattice site $j$ we then assign two SU($N_c$) matrices $V_{(A),j}$ and $V_{(B),j}$, 
each of which defines a pure gauge configuration with the link variables
\begin{equation}\label{eq:links}
  U^{i}_{A(B),j} = V_{A(B),j}V^\dag_{A(B),j+\hat{e_i}}\,,
\end{equation}
where $+\hat{e_i}$ indicates a shift from $j$ by one lattice site in the $i=1,2$ transverse direction. 
Note that Eq.\,(\ref{eq:links}) is a gauge transform of the unit matrix, hence a gauge transform of the vacuum $A^i=0$.

The link variables in the future light-cone $U_{j}^{i}$, are 
determined from solutions of the lattice CYM equations at $\tau=0$,
\begin{align}
  &{\rm tr} \left\{ t^a \left[\left(U^{i}_{(A)}+U^{i}_{(B)}\right)(1+U^{i\dag})\right.\right.\nonumber\\
   &  ~~~~~~~~~ \left.\left.-(1+U^{i})\left(U^{i\dag}_{(A)}+U^{i\dag}_{(B)}\right)\right]\right\}=0\,,\label{eq:initU}
\end{align}
where $t^a$ are the generators of $SU(N_c)$ in the fundamental representation (The cell index $j$ is omitted here).
Eq.\,(\ref{eq:initU}) reduces to Eq.\,(\ref{eq:init1}) in the continuum limit, which can be shown by expanding all
links for small $a$: $U^{i}\approx 1+iagA^{i}$.
The $N_c^2-1$ equations\,(\ref{eq:initU}) are highly non-linear and for $N_c=3$ are solved iteratively. 

We further need the lattice expression corresponding to Eq.\,(\ref{eq:init2}), which is the longitudinal electric field in the forward
light-cone \cite{Krasnitz:1998ns}
\begin{align}
  E_\eta&(\xt) = \nonumber \\
  &\frac{i}{4g} \sum_{i=x,y} \Big[
      \left( U_{(A)}^i(\xt)-U_{(B)}^i(\xt) \right) \left( U^{i\dag}(\xt) - 1 \right) \nonumber\\
  &~~~~~~~~~~~~ - h.c. -\left(U^{i\dag}(\xt-\mathbf{i}_T)-1\right) \nonumber\\
  &~~~~~~~~~~~~ \times \left(U^i_{(A)}(\xt-\mathbf{i}_T)-U^i_{(B)}(\xt-\mathbf{i}_T)\right) \nonumber \\
  &~~~~~~~~~~~~ + h.c. \Big]\,,
\end{align}
where we have indicated the cell by its coordinate $\xt$ instead of $j$ for clarity. $-\mathbf{i}_T$ indicates the shift in the $-i$ direction by one lattice spacing.

The lattice Hamiltonian in the boost invariant case is given by \cite{Krasnitz:1998ns,Lappi:2003bi}
\begin{align}\label{eq:ham}
  aH &= \sum_{\xt} \left[ \frac{g^2 a}{\tau} {\rm tr}\, E^i E^i + \frac{2 \tau}{g^2 a}\left(N_c-{\rm Re}\,{\rm tr}\, U_{1,2}\right)\right.\nonumber\\
   \left.\right. &~~~~~~~~~~~\left.+\frac{\tau}{a}{\rm tr}\, \pi^2 + \frac{a}{\tau}\sum_i{\rm tr} \left(\phi - \tilde{\phi}_i\right)^2\right]\,,
\end{align}
where the sum is over all cells in the transverse plane. For clarity, we have omitted the cell index $j$ for all quantities in this expression.
$\phi$ is a scalar field, resulting from $A_\eta$ when disallowing $\eta$ dependent gauge transformations, and $\pi=E_\eta=\dot{\phi}/\tau$ is the
longitudinal electric field. $E^i$ with $i\in\{1,2\}$ are the components of the transverse electric field that initially are zero.
The parallel transported scalar field in cell $j$ is given by
\begin{equation}
  \tilde{\phi}_{i}^j = U_j^i \phi_{j+\hat{e}_i} U_j^{i\dag}\,,
\end{equation}
and the plaquette is given by
\begin{equation}\label{eq:plaq}
 U_{1,2}^j = U^1_j \, U^2_{j+\hat{e}_1} \, U^{1\dag}_{j+\hat{e}_2} \, U^{2\dag}_j\,.
\end{equation}

With only longitudinal fields present after the collision, the total energy density on the lattice at $\tau=0$ is given by
\begin{equation}\label{eq:eden}
  \varepsilon(\tau=0) = \frac{2}{g^2 a^4} (N_c - {\rm Re}\,{\rm tr}\, U_{1,2}) + \frac{1}{a^4} {\rm tr}\,\pi^2\,,
\end{equation}
where the first term is the longitudinal magnetic, the second the longitudinal electric energy density.

To evolve the system from $\tau = 0$ forward, we use the Hamiltonian equations of motion obtained from taking the Poisson brackets of the fields
with the Hamiltonian (\ref{eq:ham})
\begin{align}
  \dot{U}_i &= i\frac{g^2}{\tau}E^i U_i ~ ({\rm no~sum~over~} i)  \\
  \dot{\phi} &= \tau \pi \\
  \dot{E}^1 &= \frac{i\tau}{2 g^2} [U_{1,2} + U_{1,-2} - U_{1,2}^\dag - U_{1,-2}^\dag - T_1] \nonumber\\
  & ~~~+ \frac{i}{\tau}[\tilde{\phi_1},\phi] \\
  \dot{E}^2 &= \frac{i\tau}{2 g^2} [U_{2,1} + U_{2,-1} - U_{2,1}^\dag - U_{2,-1}^\dag - T_2] \nonumber\\
  & ~~~+ \frac{i}{\tau}[\tilde{\phi_1},\phi]\\
  \dot{\pi} &= \frac{1}{\tau} \sum_i \left[\tilde{\phi}_i+\tilde{\phi}_{-i} - 2\phi\right]
\end{align}
where $T_1 = \frac{1}{N_c}{\rm tr}[U_{1,2} + U_{1,-2} - U_{1,2}^\dag - U_{1,-2}^\dag]~ {\mathbf 1}$, 
and $T_2 = \frac{1}{N_c}{\rm tr}[[U_{2,1} + U_{2,-1} - U_{2,1}^\dag - U_{2,-1}^\dag]~ {\mathbf 1}$ with the $N_c \times N_c$ unit matrix ${\mathbf 1}$.
The subtraction of these traces takes care of the fact that the components of $F^{\mu\nu}$ are traceless color matrices.
Note that the Hermitian conjugates of a plaquette simply reverse direction.

To determine the energy density at times $\tau>0$ we evaluate the $\tau\tau$-component of the energy-momentum tensor
at site $j$ 
\begin{align}\label{eq:Ttt}
  T^{\tau\tau}_j &= \frac{g^2}{2\tau^2} {\rm tr} \left[ E_{x, j}^2 + E_{x, j+\hat{e}_2}^2 + E_{y, j}^2 + E_{y, j+\hat{e}_1}^2\right]\nonumber \\
  & +\frac{1}{2\tau^2}{\rm tr} \Big[(\phi_j-U_{x, j} \phi_{j+\hat{e}_1} U^\dag_{x, j})^2\nonumber\\
  & ~~~~~~~~~~~~ +(\phi_{j-\hat{e}_1}-U_{x, j-\hat{e}_1} \phi_{j} U^\dag_{x, j-\hat{e}_1})^2\Big]\nonumber\\
  & +\frac{1}{2\tau^2}{\rm tr} \Big[(\phi_j-U_{y, j} \phi_{j+\hat{e}_2} U^\dag_{y, j})^2\nonumber\\
  & ~~~~~~~~~~~~ +(\phi_{j-\hat{e}_2}-U_{y, j-\hat{e}_2} \phi_{j} U^\dag_{y, j-\hat{e}_2})^2\Big]\nonumber\\
  & + {\rm tr} \big[ \pi_j^2 \big] + \sum_{4\square}\frac{2}{g^2}\Big(N_c-{\rm Re}\,{\rm tr}\big[U_\square\big]\Big)\,,
\end{align}
where $\sum_{4\square}$ indicates the sum over the plaquette defined in (\ref{eq:plaq}) 
and the three other plaquettes with the same orientation connected to site $j$.
Note that care has to be taken that all quantities are defined at the same point, in this case lattice site $j$.


\section{Results}
\subsection{Energy density and $dN_g/dy$}
\begin{figure}[h]
  \includegraphics[width=8cm]{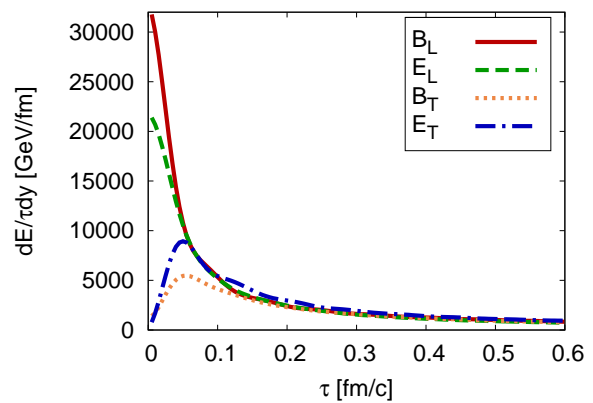}
  \caption{Components of the integrated energy density $dE/\tau dy$ as a function of $\tau$ from one single event. Labels indicate components of
     $dE/\tau dy$.
    \label{fig:energy}}
\end{figure}

Having outlined the numerical procedure, we will now present results from the CYM evolution of the IP-Sat initial conditions. We begin by showing the evolution of the energy density (\ref{eq:Ttt}), integrated over the transverse plane $\frac{dE}{\tau dy}$ in Fig.\,\ref{fig:energy}.
As can be seen from Eq.\,(\ref{eq:eden}), initially the system consists of only longitudinal chromo-magnetic and chromo-electric fields. Over the course of
the evolution, the transverse components of the energy density grow and catch up with the longitudinal modes.
The expansion of the system leads $dE/\tau dy$ to drop like $1/\tau$ and $dE/dy$ becomes a constant.

\begin{figure}[tb]
   \begin{center}
     \includegraphics[width=7cm, clip, trim=3.5cm 0cm 0cm 1cm]{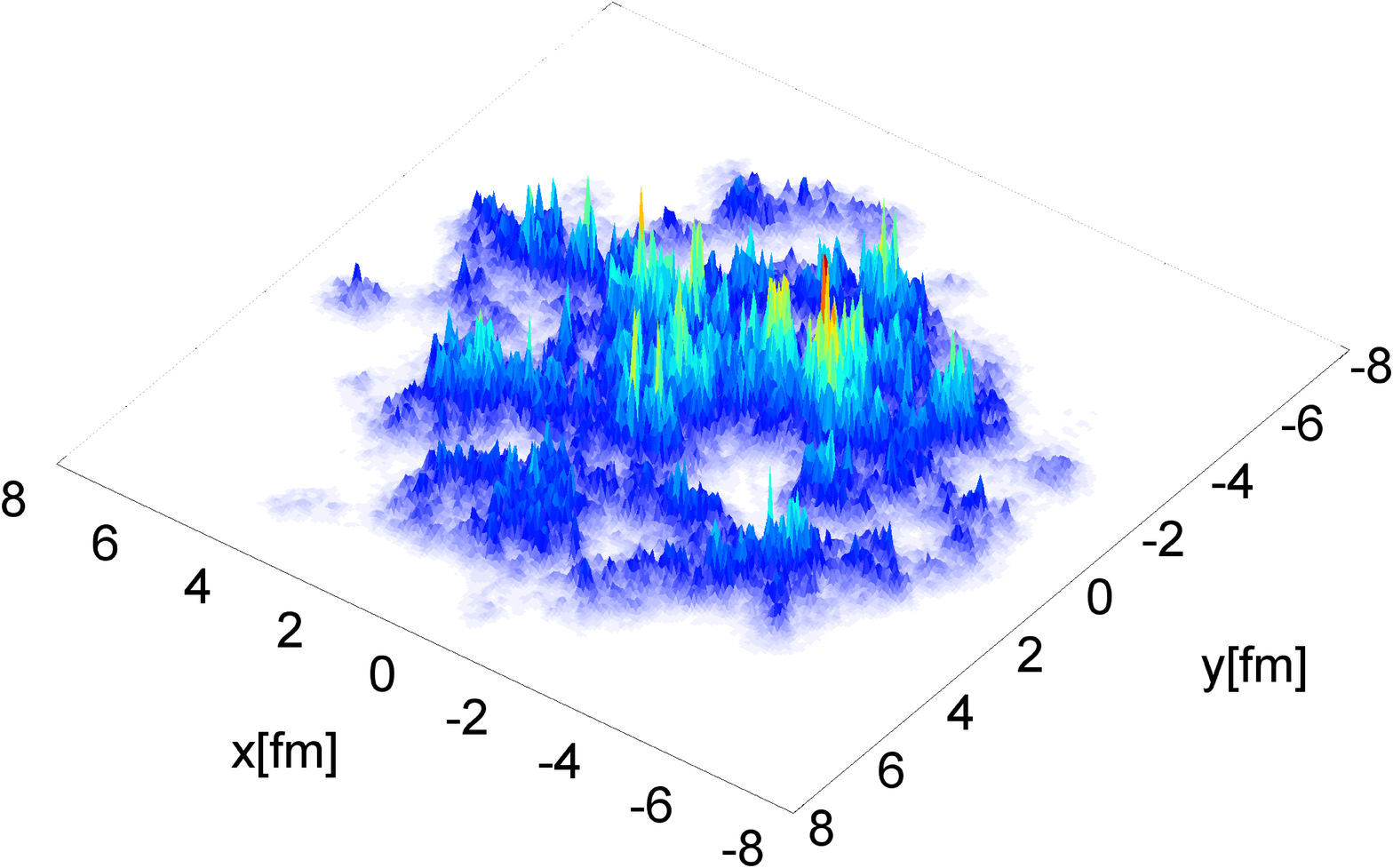}\\\vspace{-0.5cm}
     \includegraphics[width=7cm, clip, trim=3.5cm 0cm 0cm 1cm]{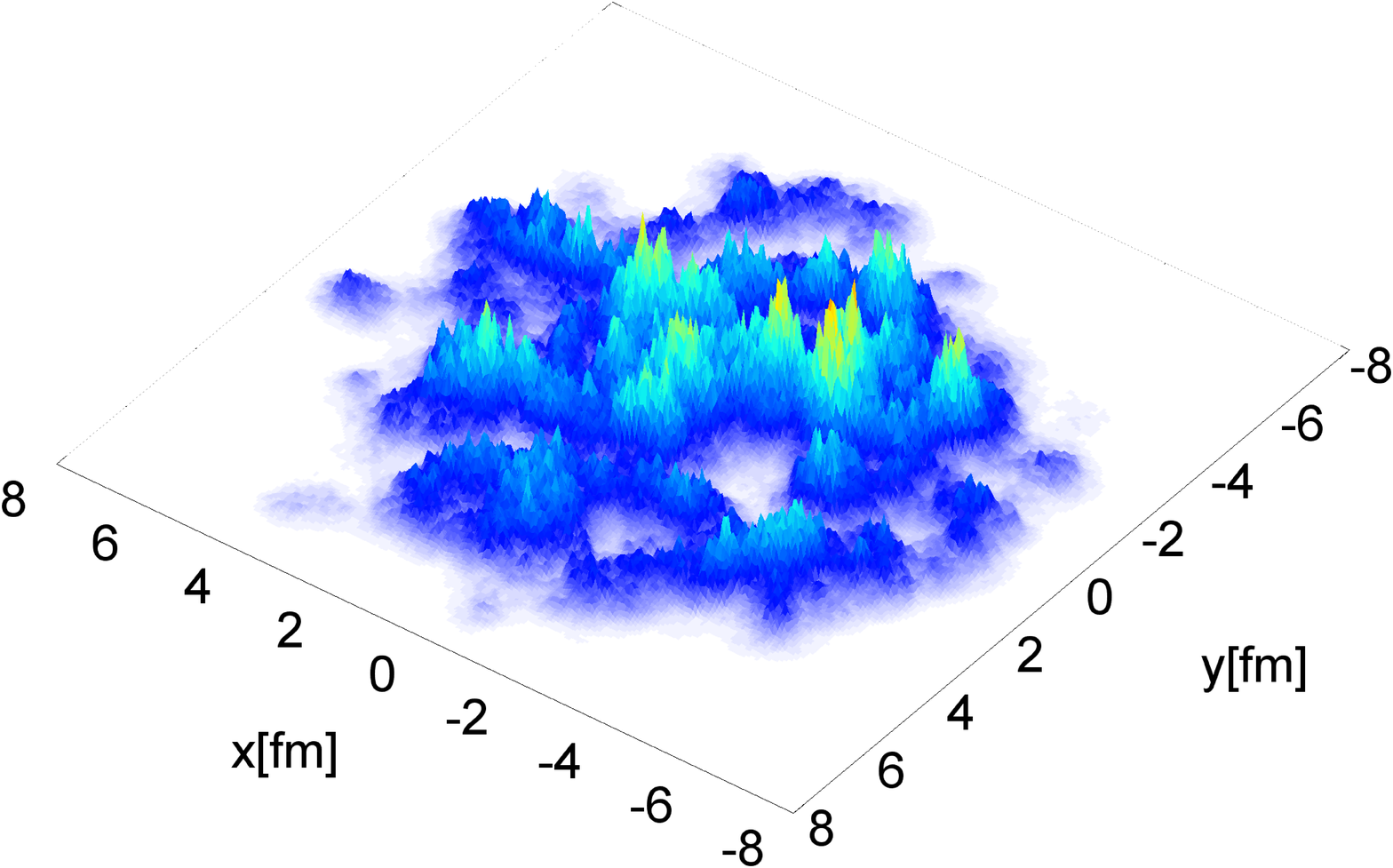}
     \vspace{-0.5cm}
     \caption{(Color online) Energy density (arbitrary units) in the transverse plane at $\tau=0\,{\rm fm}$ (upper panel) 
       and $\tau=0.2\,{\rm fm}$ (lower panel). The structures are smoothed by the evolution over the first $\Delta\tau\sim 1/Q_s$. }
     \label{fig:eps}
   \end{center}
\vspace{-0.7cm}
\end{figure}

Fig.\,\ref{fig:eps} shows the structure of the energy density in the transverse plane after the collision, at time $\tau=0\,{\rm fm}$ (Eq.\,(\ref{eq:eden})) and after classical Yang-Mills evolution in 2+1 dimensions for $\Delta\tau = 0.2\,\,{\rm fm}$, which is of the order of $1/Q_s$, the time scale over which
interactions are still significant. After this time, expansion causes the fields to become weak and the system becomes freely streaming.
As shown in Fig.\,\ref{fig:energy}, at the later time all four components of Eq.\,(\ref{eq:ham}) contribute. 
As can be expected, the distribution becomes smoother with the evolution.

We next compute the gluon multiplicity per unit rapidity 
$dN_g/dy$. Since the multiplicity is not a gauge invariant quantity, we fix transverse Coulomb gauge ($\partial_i A^i=0$, with $i$ summed over $1,2$). Then the lattice
expression for $dN_g/dy$ is given by \cite{Krasnitz:2001qu,Lappi:2003bi}
\begin{align}
  \frac{dN_g}{dy} = \frac{2}{N^2} \int \frac{d^2k_T}{\tilde{k}_T} &\Big[\frac{g^2}{\tau} {\rm tr} \left(E_i(\kt) E_i(-\kt)\right)\nonumber\\
    & ~~ + \tau\, {\rm tr}
    \left(\pi(\kt)\pi(-\kt)\right)\Big]\,,
\end{align}
assuming a free massless lattice dispersion relation for the interacting theory. This leads to the appearance of the square root of
\begin{equation}
  \tilde{k}_T^2 = 4 \left[\sin^2\frac{k_x}{2}+\sin^2\frac{k_y}{2}\right]\,.
\end{equation}

\begin{figure}[h]
 \includegraphics[width=8cm]{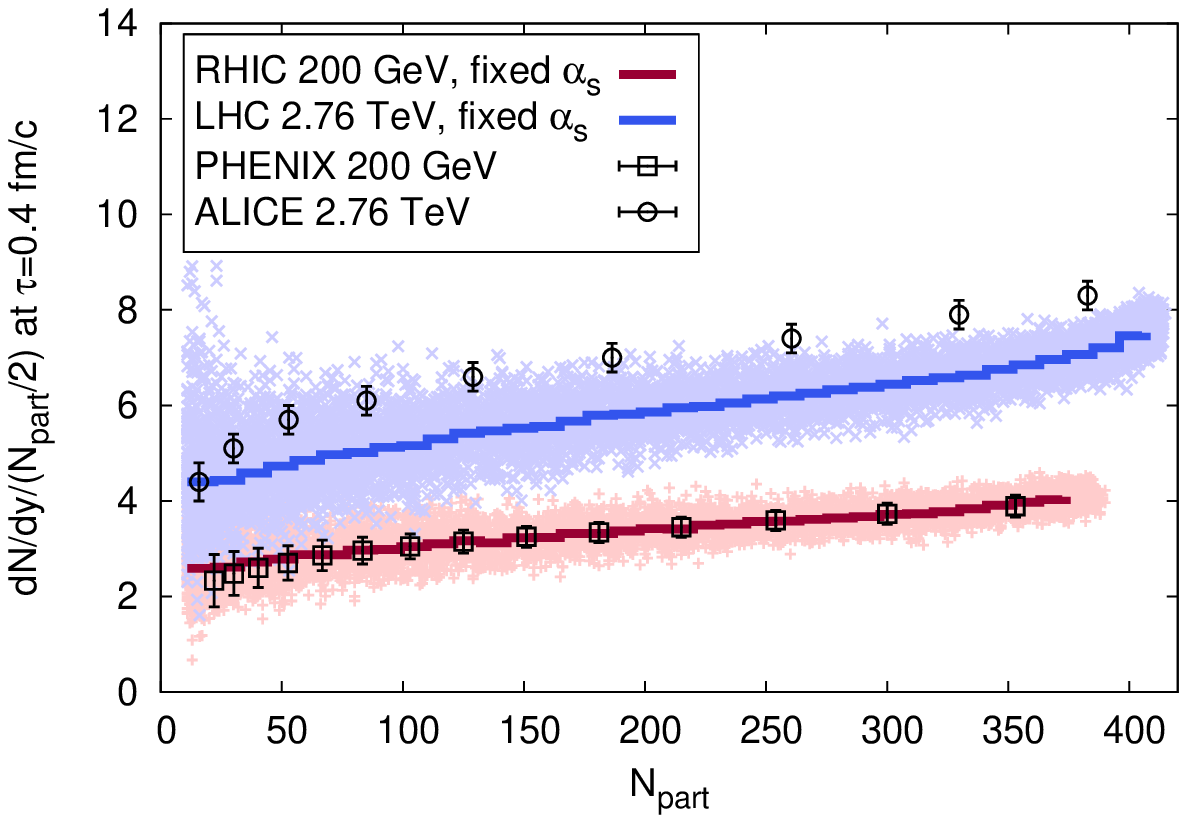}\\
 \includegraphics[width=8cm]{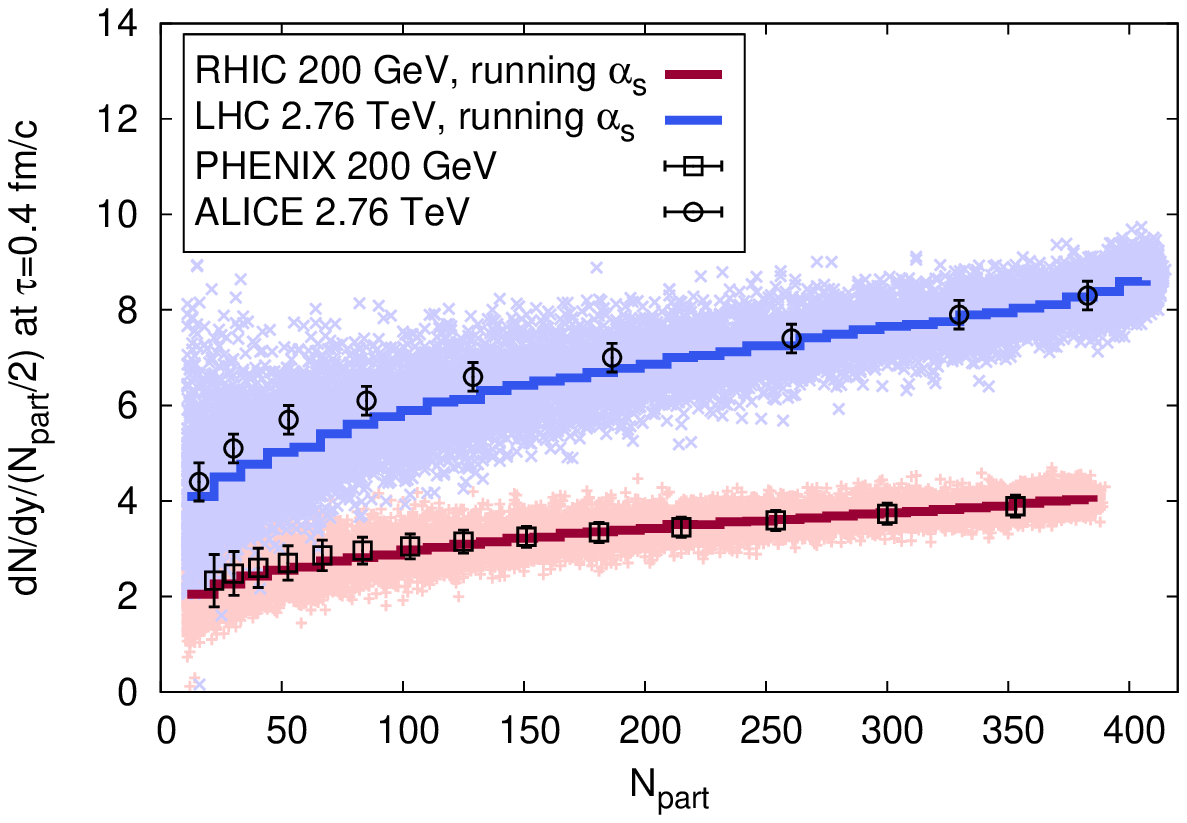}
  \caption{Gluon multiplicity $(dN_g/dy)/(N_{\rm part}/2)$ at $\tau=0.4\,{\rm fm}/c$ times $2/3$ compared to experimental 
    charged particle $(dN/dy)/(N_{\rm part}/2)$ data for $\sqrt{s}=200\,{\rm GeV}$ Au+Au
    and $\sqrt{s}=2.76\,{\rm TeV}$ Pb+Pb collisions as a function of $N_{\rm part}$ for
    fixed coupling (upper panel) and running coupling (lower panel). 
    The pale blue and red bands are a collection of the multiplicities for individual events, 
    with the solid lines representing the average multiplicity.  Experimental data from \cite{Adler:2004zn} and \cite{Aamodt:2010cz}. 
    \label{fig:dNdyNpart}}
\end{figure}

The result for $(dN_g/dy)/(N_{\rm part}/2)$ vs. $N_{\rm part}$ at time $\tau=0.4\,{\rm fm}$, where $N_{\rm part}$ is the number of participant nucleons, is shown in Fig.\,\ref{fig:dNdyNpart}. The IP-Glasma  model does not have the concept of ``wounded nucleons" because we treat the nucleus as a coherent 
system of gluon fields correlated on distance scales $1/Q_s$ much smaller than the size of a nucleon.  To determine $N_{\rm part}$ we use the same  Monte Carlo Glauber (MC-Glauber) model as employed by the experimental collaborations. 
Nucleons that were sampled for each nucleus, as described in Section \ref{sec:numcalc},
are assumed to be participant nucleons if the relative transverse distance between them and a nucleon from the other nucleus 
is smaller than $D=\sqrt{\sigma_{NN}/\pi}$, where $\sigma_{NN}$ is the 
total inelastic cross section, $\sigma_{NN}=42\,{\rm mb}$ for $\sqrt{s}=200\,{\rm GeV}$ 
Au+Au collisions at RHIC and $64\,{\rm mb}$ for $\sqrt{s}=2.76\,{\rm TeV}$ Pb+Pb collisions at LHC. The reader should note  that the MC-Glauber model
and the inelastic cross sections are solely used to determine $N_{\rm part}$ to compare to experimental data. 
They are not an input for the IP-Glasma model.

In the upper panel of Fig.\,\ref{fig:dNdyNpart}, we show the multiplicity distribution as a function of $N_{\rm part}$ for fixed coupling. We multiplied the gluon multiplicity by $2/3$ to convert
to charged particle multiplicity. Note that the overall normalization is chosen (by varying the ratio between $Q_s$ and 
$g^2\mu$ and $\alpha_s$ in the fixed coupling case) to agree with the RHIC data for charged particles. This allows us to better compare the shape of the result and the experimental data. 
(The pale bands denote results from the individual events and demonstrate the range of 
fluctuations around the mean. See below for more details.) 
However, we know that there is entropy production in the system and the initial gluon multiplicity should not account for all observed final particles. The logarithmic uncertainty in $Q_s$ as well as some numerical uncertainty (for details see \cite{Lappi:2007ku,Lappi:2009xa}) in the factor between $Q_s$ and $g^2\mu$ introduce some freedom that allows to adjust the normalization of the initial 
$dN_g/dy$. This also allows to adjust the energy density when fine tuning to experimental data when using this model with a viscous hydrodynamic evolution model that accounts for entropy production.

While the RHIC result is reasonably well described, both the normalization and shape of the LHC result disagree strongly with the experimental data for the fixed coupling case. Note that the IP-Sat model naturally generates different $x$ dependencies of $Q_s$ for protons and nuclei, the latter giving a stronger dependence on $x$. Because we use the parametrization of $Q_s$ from DIS off of protons to construct nuclei, we get the right spatial dependence of color charge distributions but the $x$ dependence for nuclei is off, and the stronger $x$-dependence  has to be introduced by hand. We observe that our results are consistent with Ref.~\cite{Tribedy:2011aa}. 

The inclusion of running coupling as shown in the lower panel of Fig.\,\ref{fig:dNdyNpart}
improves the agreement with the LHC data significantly. The shape is now very well described apart from a very modest  under-prediction at small $N_{\rm part}$. This discrepancy could be from entropy production in the later evolution stages, whose relative contribution can be expected to be somewhat larger for smaller systems.
The scale $\tilde{\mu}$ for the running of 
\begin{equation}\label{eq:runningalpha}
  \alpha_s(\tilde{\mu}) = \frac{2\pi}{(11-\frac{2}{3}N_f) \ln (\tilde{\mu}/\Lambda_{\rm QCD})}\,,  
\end{equation}
was chosen to be $\tilde{\mu}=\langle{\rm max}(Q_s^A(\xt),Q_s^B(\xt))\rangle$, where the average is over the whole transverse
plane. As usual, there is some ambiguity in choosing the scale, but we have verified that using the minimum instead of the maximum or twice the value did not make a significant difference for the shape of the result. Note that the CYM evolution does not depend on $\alpha_s$ as it scales out from the equations of motion and only enters in the final computation of the multiplicity. Furthermore note that the use of MC-Glauber to determine $N_{\rm part}$ is crucial for reproducing the experimentally observed shape of the result. In this way, we include both fluctuations in 
$N_{\rm part}$ (along the horizontal axis) as well as in $dN/dy$ (along the vertical axis). Individual events (15000 each) are plotted as scattered points in the background to show the degree of fluctuations explicitly.
In a forthcoming paper, we will evaluate how event-by-event viscous hydrodynamic simulations modify this computation.

\subsection{Multiplicity distributions}
\begin{figure}[h]
  \includegraphics[width=8cm]{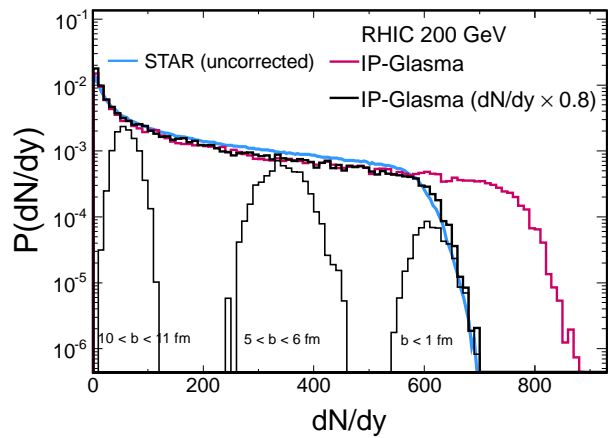}
  \caption{Probability distribution of gluon multiplicities $dN/dy$ at $\tau=0.4\,{\rm fm}/c$.
    Shown are also the distributions for some limited ranges of impact parameter $b$, which are described by negative binomial distributions.
    Experimental data from STAR \cite{Abelev:2008ez}.
    \label{fig:dNdyweighted}}
\end{figure}

In Fig.\,\ref{fig:dNdyweighted} we present the probability distribution of $dN_g/dy$ at RHIC energies. An essential ingredient is the probability distribution of impact parameters, which is determined by the Glauber model to be
\begin{equation}
  \frac{dP_{\rm inel}}{d^2\bt} = \frac{1-(1-\sigma_{NN} T_{AB})^{AB}}{\int d^2\bt \left(1-(1-\sigma_{NN} T_{AB})^{AB}\right)}\,,
\end{equation}
with the overlap function $T_{AB}$. One could in principle compute this distribution in the Glasma framework, but a first principles computation is extremely difficult. The probability for no inelastic interaction (an essential ingredient in the above equation) at a given impact parameter requires an understanding of diffractive/elastic interactions which is incomplete at present in all QCD based frameworks. The Glauber model, with parameters tuned to data, is therefore a good effective model for this aspect of our computation. 

We compute the n-particle multiplicity distribution by first sampling the impact parameter $b$ from a uniform distribution, computing the resulting $dN_g/dy$ from the IP-Glasma model,
and when binning these values into the histogram shown in Fig.\,\ref{fig:dNdyweighted}, weight the result with a factor $2\pi\, b\, dP_{\rm inel}/d^2\bt$ depending on the $b$ value used in a given event.
The STAR data shown is uncorrected, which makes a direct comparison difficult. We therefore scale $dN/dy$ by a factor of 0.8 and achieve good 
agreement with the data. Note, however, that the scaling factor should depend on $dN/dy$ \cite{Abelev:2008ez}.

\begin{figure}[h]
  \includegraphics[width=8cm]{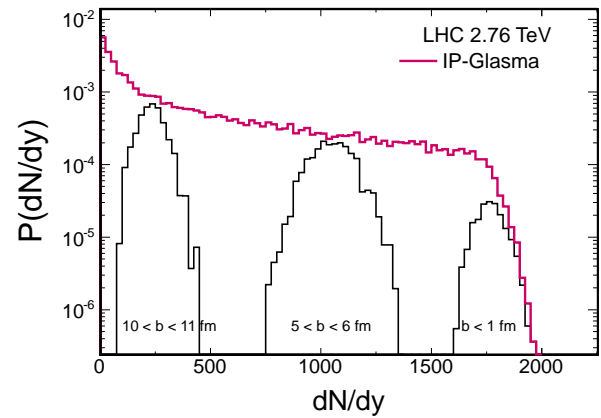}
  \caption{Same as Fig. \ref{fig:dNdyweighted} for LHC energy $\sqrt{s}=2.76\,{\rm TeV}$.\label{fig:dNdyweightedLHC}}
\end{figure}

We also show three distributions obtained by constraining the impact parameter range to demonstrate that their shape resembles a negative binomial
distribution (NBD). The negative binomial fluctuations of the transverse energy as emerging from the IP-Glasma model have previously been discussed in \cite{Schenke:2012wb}. Here we show that they also appear in the multiplicity distribution.
In the Glasma flux tube framework \cite{Dumitru:2008wn,Gavin:2008ev}, negative binomial distributions
\begin{equation}\label{eq:nbd}
  P_n^{\rm NB}(\bar{n},k) = \frac{\Gamma(k+n)}{\Gamma(k)\Gamma(n+1)} \frac{\bar{n}^n k^k}{(\bar{n}+k)^{n+k}}\,,
\end{equation}
with
\begin{equation}\label{eq:k}
  k = \zeta \frac{N_c^2-1}{2\pi} Q_s^2 S_\perp
\end{equation}
arise~\cite{Gelis:2009wh}, with  $k$ inversely proportional to the width of the NBD. $k$ here is proportional to the number of flux tubes $Q_s^2 S_\perp$, where
$S_\perp$ is the transverse size of the system. (Hence smaller systems at a fixed energy generate more fluctuations.) 
In the expression above, $\zeta$ is an intrinsically non-perturbative function which can be computed {\it ab initio} from solutions of the CYM equations. 
To determine $\zeta$ it is sufficient to compute the double inclusive distribution by solving CYM equations as done in Ref.~\cite{Lappi:2009xa}.
Here we are computing the n-particle inclusive distribution, and can thus also extract $\zeta$ if the NBD description of these distributions is robust. $\zeta$ in general can depend on $Q_s^2 S_\perp$; however, a powerful test of how robust the Glasma flux tube picture is depends on this dependence being weak for large values of $Q_s^2 S_\perp$. We will therefore discuss the behavior of $\zeta$ below and what the results tell us about the nature of different sources of quantum fluctuations.

Within the IP-Glasma model, we determine $\zeta(Q_s^2 S_\perp)$ by extracting $k$ from a fit 
with an NBD and computing an average $\langle Q_s^2 S_\perp\rangle$ by summing over the minimum of the two $Q_{s, A(B)}^{2}$ in the whole transverse plane. 

\begin{figure}[h]
  \includegraphics[width=8.5cm]{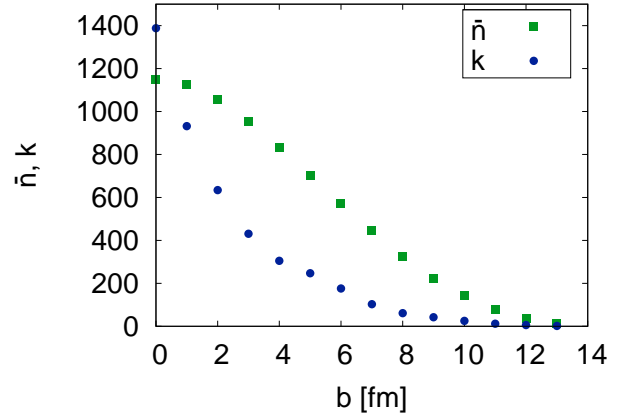}
  \caption{Average gluon number $\bar{n}$ and width parameter $k$ as a function of impact parameter $b$ at $\sqrt{s}=200\,{\rm GeV}$. \label{fig:nbark}}
\end{figure}

We first determine $\bar{n}$ and $k$ of Eq.\,(\ref{eq:nbd}) from the fit to the multiplicity distributions at fixed impact parameter $b$. The results are shown in Fig.\,\ref{fig:nbark}.
Interestingly, the ratio $k/\bar{n}$ is greater than one for central collisions and at large impact parameters 
approaches approximately $k/\bar{n}\approx 0.14$, which is close to the value determined from fits
to distributions in p+p collisions \cite{Dumitru:2012yr}.

\begin{figure}[h]
  \includegraphics[width=8.5cm]{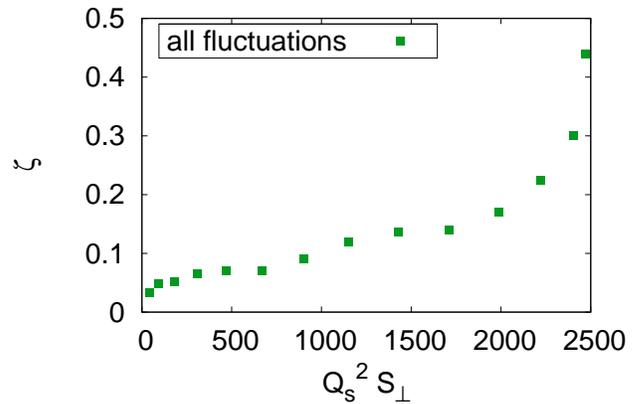}
  \caption{Proportionality factor $\zeta$ at RHIC energy in $k = (\zeta (N_c^2-1)/(2\pi)) Q_s^2 S_\perp$ as a function of $Q_s^2S_\perp$ 
    including all fluctuations. $Q_s^2S_\perp$ corresponds to the average value for a given $b$. $b$ was varied in $1\,{\rm fm}$ steps
    from $0$ to $13\,{\rm fm}$. \label{fig:zetab}}
\end{figure}

The corresponding $\zeta$-values are shown in Fig.\ref{fig:zetab} as a function of the average 
values of $Q_s^2 S_\perp$ for a given $b$. We observe a strong dependence of 
$\zeta$ on $Q_s^2 S_\perp$, which is in disagreement with the flux tube picture. The reason is that the 
effect of fluctuations in the number of wounded nucleons (which were not considered in the derivation of Eq.\,(\ref{eq:k})) 
in addition to fluctuations in the color charge distributions,
make the distribution wider ($\zeta$ smaller), especially at large impact parameters (small $Q_s^2 S_\perp$) where geometrical fluctuations dominate.

This behavior of $\zeta$ is compatible with previously extracted values from fits based on $k_\perp$ factorization to A+A multiplicity distributions~\cite{Tribedy:2011aa}, where small $dN/dy$ required small values of $zeta$ and large $dN/dy$ larger values to achieve a good fit. The IP-Glasma model 
automatically produces this variation of $\zeta$, leading to very good agreement with the experimental data as shown in Fig.\,\ref{fig:dNdyweighted}.

To be able to better compare to the flux tube picture and Eq.\,(\ref{eq:k}), we now consider only the effect of fluctuations in color charges.
To do so we average over the nuclear color charge density squared over many nucleon configurations. This results in a smooth distribution that removes fluctuations in the wounded nucleon number and positions.

\begin{figure}[h]
  \includegraphics[width=8cm]{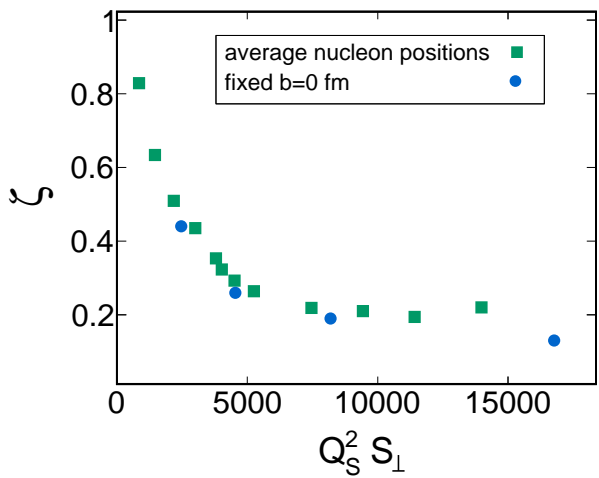}
  \caption{Proportionality factor $\zeta$ in $k = (\zeta (N_c^2-1)/(2\pi)) Q_s^2 S_\perp$ as a function of $Q_s^2 S_\perp$ for averaged nucleon positions (squares) and with nucleon fluctuations at fixed impact parameter $b=0\,{\rm fm}$ (circles). 
    At large $Q_s^2 S_\perp$ the result for the smooth distribution approaches a constant as predicted by the Glasma flux tube model for n-gluon correlations. The result for fluctuating nucleon positions at constant $b=0\,{\rm fm}$ is very similar and becomes very weakly dependent on $Q_s^2S_\perp$.
    \label{fig:zeta}}
\end{figure}

This is in the spirit of the original Glasma flux tube perturbative~\cite{Gelis:2009wh} and non-perturbative~\cite{Lappi:2009xa} computations. The result
for this $\zeta_{\rm smooth}$ is shown in Fig.\,\ref{fig:zeta}.
After starting out near 1, $\zeta_{\rm smooth}$ drops and approaches a constant value of approximately $\zeta_{\rm smooth}=0.2$ for large $Q_s^2 S_\perp$.
This means that at low parton densities, $k$ is initially more constant than expected in the Glasma flux tube picture but then becomes proportional to $Q_s^2 S_\perp$ at high parton densities as anticipated.

The fact that $\zeta$ at small impact parameters (cf. Fig.\,\ref{fig:zetab}) approaches that in the smooth case for the same $Q_s^2 S_\perp$ gives hope that
there is a chance to experimentally access a regime where the flux tube picture is valid. Fixing $b=0\,{\rm fm}$ and increasing the energy,
and this way increasing $Q_s^2 S_\perp$ while keeping $S_\perp$ as constant as possible, reduces fluctuations in the nucleon number.
Indeed we find that the result for the extracted $\zeta$, shown as blue circles in Fig.\,\ref{fig:zeta},
is very close to the one obtained with smooth initial distributions, and its dependence on $Q_s^2S_\perp$ becomes weak at large $Q_s^2S_\perp$.

\subsection{Eccentricities}
\begin{figure}[h]
  \includegraphics[width=7.5cm]{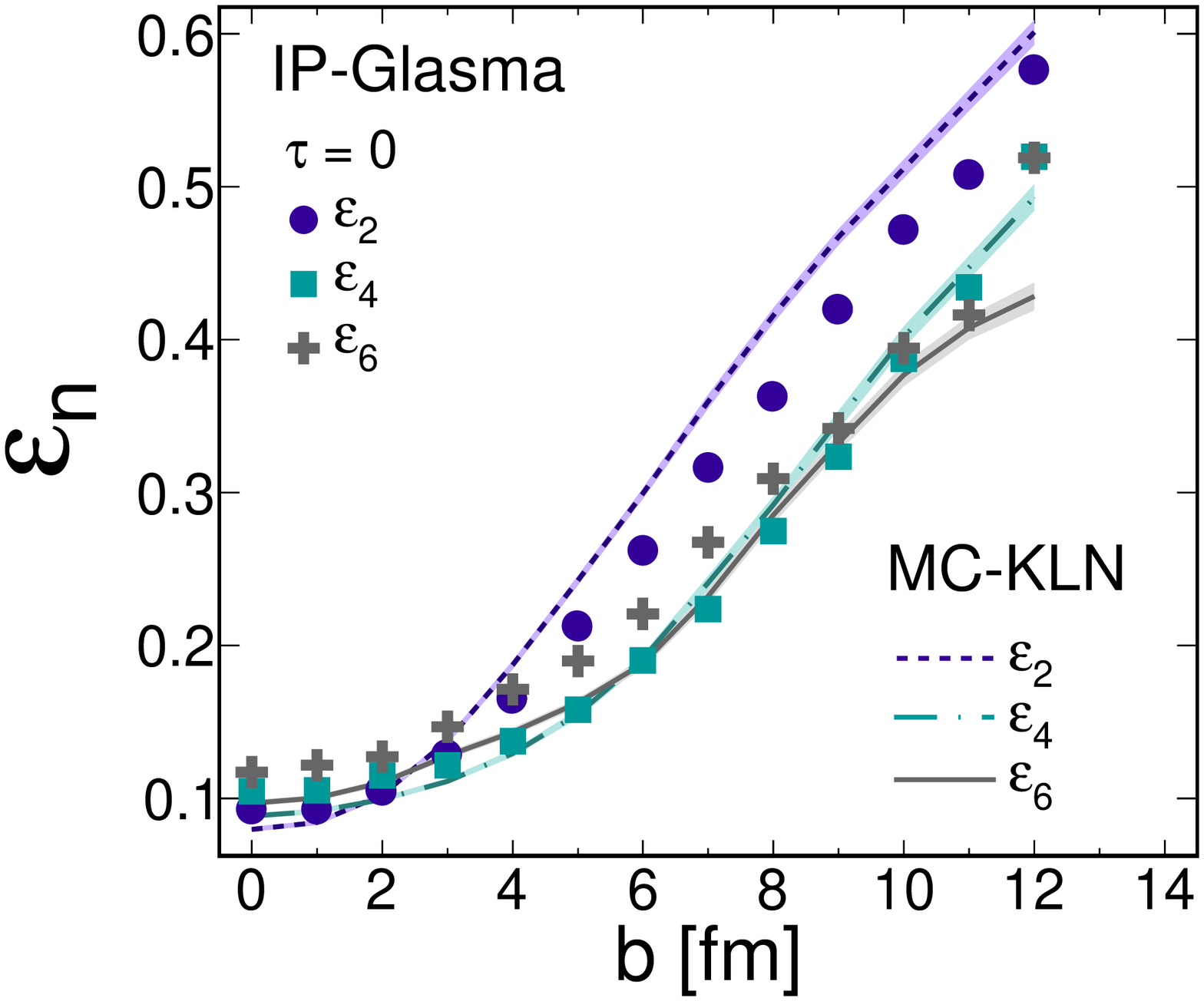}\\
  \includegraphics[width=7.5cm]{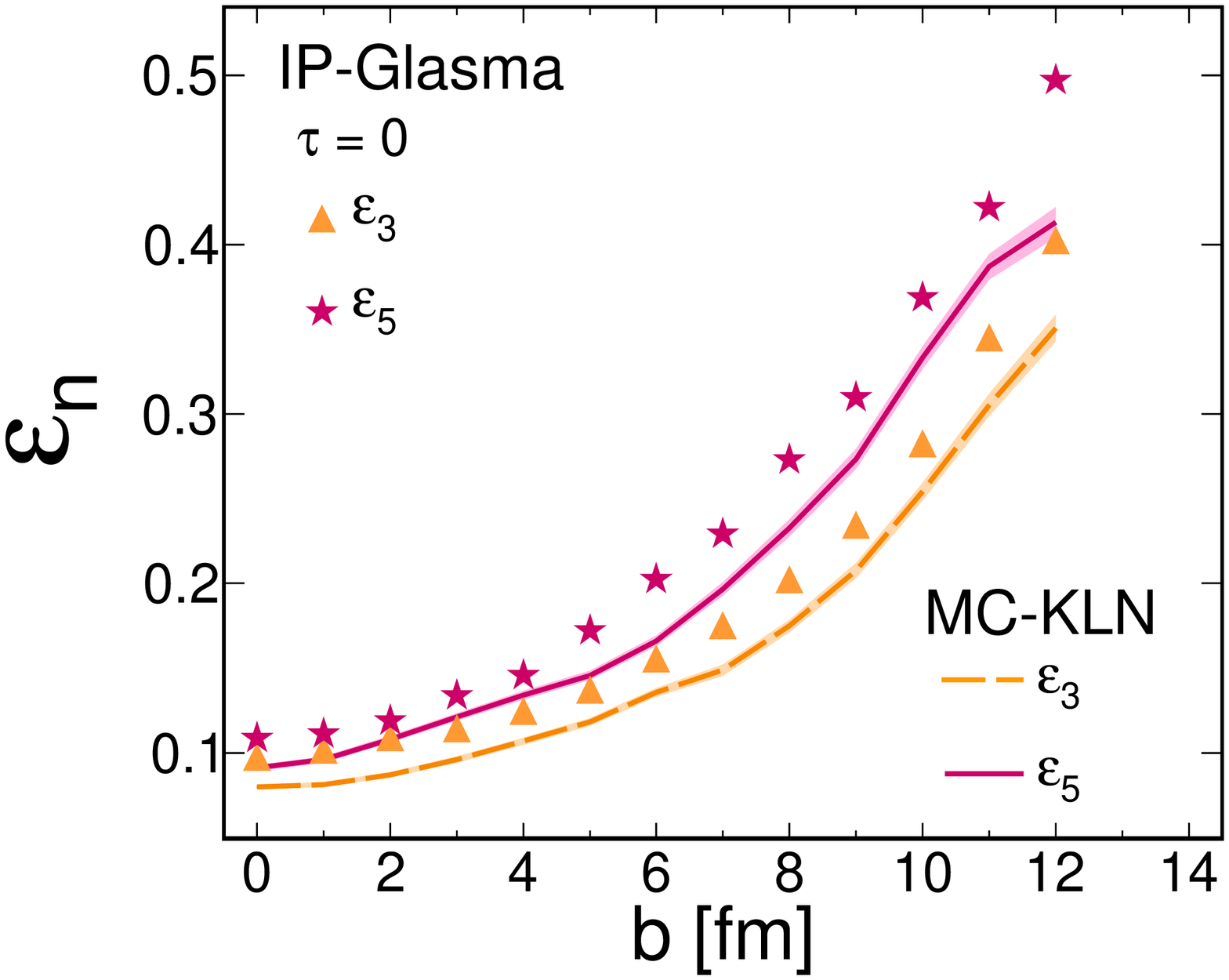}
  \caption{Even (upper panel) and odd (lower panel) eccentricities from the IP-Glasma model compared to those from MC-KLN.
    \label{fig:ecc}}
\end{figure}

\begin{figure}[h]
  \includegraphics[width=8.5cm]{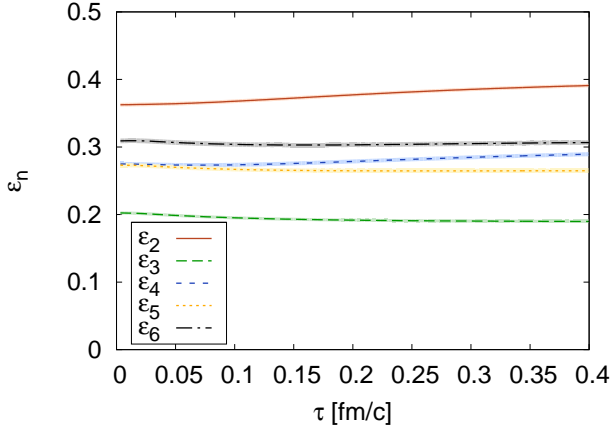}
  \caption{Time evolution of the eccentricities from the IP-Glasma model at impact parameter $b=8\,{\rm fm}$.
    \label{fig:ecctime}}
\end{figure}

In Ref.\,\cite{Schenke:2012wb}, we presented results for $\varepsilon_2$ and $\varepsilon_3$ defined in (\ref{eq:eccen}) and compared to
results from an MC-Glauber model and the MC-KLN model \cite{Hirano:2005xf,Drescher:2006pi,*Drescher:2007ax}. Here we extend this study 
by comparing eccentricities up to $\varepsilon_6$.
The results are shown in Fig.\,\ref{fig:ecc} and the conclusions that can be drawn are mainly that the purely fluctuation driven odd harmonics
$\varepsilon_3$ and $\varepsilon_5$ from the IP-Glasma model are larger than those from the MC-KLN model \cite{mckln} for all $b$, while $\varepsilon_2$
is smaller than that computed in the MC-KLN model, in particular for $b>3\,{\rm fm}$.
As a consequence, the ratio $\varepsilon_2/\varepsilon_3$ is smaller than in the MC-KLN model, which is going to decrease the ratio of $v_2/v_3$ obtained after hydrodynamic evolution, making it more compatible with experimental observation. 
$\varepsilon_4$ and $\varepsilon_6$ are almost equal or larger than those from the MC-KLN model. 
We make the comparison to MC-KLN at $\tau=0\,{\rm fm}/c$, because at later times we would also have to take into account the pre-equilibrium
flow built up in the CYM simulation. 
This effect will be included in detailed event-by-event simulations that convert the spatial anisotropies into momentum anisotropies
in a follow up paper in this series.
To show the effect of the CYM evolution on the eccentricities themselves we show as an example their time evolution for $b=8\,{\rm fm}$ in Fig.\,\ref{fig:ecctime}. As expected already in \cite{Schenke:2012wb}, the change in all $\varepsilon_n$ is very weak over the first $0.4\,{\rm fm}/c$. After this time
all $\varepsilon_n$ begin dropping as the systems is freely streaming and hence becoming more isotropic.

\section{Summary and Outlook}

This paper expands on the IP-Glasma model of fluctuating initial conditions for heavy-ion collisions first presented in Ref.~\cite{Schenke:2012wb}. More details of the computations are presented as well as novel results for the single inclusive and n-gluon multiplicity distributions and higher even and odd eccentricity moments. In addition, the CYM equations are solved for finite proper times $\tau$ unlike Ref.~\cite{Schenke:2012wb} where results were extracted only for $\tau=0$. 

We observed that the running coupling CYM results give good agreement with the RHIC and LHC single inclusive multiplicity 
distributions as a function of $N_{\rm part}$. The computed n-particle inclusive multiplicity distribution shows good agreement with the uncorrected STAR data on the same once a constant correction factor is applied. A prediction is made for multiplicity distributions at the LHC. We further observe that our results are well described as a convolution of negative binomial distributions at different impact parameters. The parameters of the negative binomial distributions are extracted, and it is observed that, in the approximation of smoothed nucleon configurations, for large parton densities, the predictions of the Glasma flux tube picture are recovered. For the realistic situation of fluctuating wounded nucleon configurations, 
one still obtains NBDs albeit with significantly wider widths. The non-perturbative coefficient $\zeta$ introduced in the 
Glasma flux tube description quantifies the effect of wounded nucleon fluctuations. Predictive power, in particular for central impact parameters at higher energies,  is still retained because $\zeta$ is (nearly) energy independent, while the width parameter $k$ (in Eq.~\ref{eq:k}) has a strong energy dependence controlled by the saturation scale $Q_s^2$. 

The computation of eccentricity moments is performed up to $\epsilon_6$. It is seen that $\varepsilon_2$ is smaller than in the MC-KLN model (used as an initial condition in many hydrodynamic studies), while the odd moments are larger, pointing to the additional role of multiplicity fluctuations in the IP-Glasma model.

An essential follow up to this work is to match the results of the IP-Glasma model, event-by-event, to viscous hydrodynamic simulations. This will allow one to gauge the effects of dissipative flow in modifying the energy and multiplicity distributions and on the conversion of spatial anisotropies into momentum anisotropies. These have the potential to significantly enhance our understanding of the transport properties of the quark-gluon plasma, with the caveat that a systematic treatment of pre-equilibrium flow including instabilities can alter some of these conclusions significantly.


\section*{Acknowledgments}
BPS\ and RV\ are supported by US Department of Energy under DOE Contract No.DE-AC02-98CH10886 and acknowledge additional support from a BNL``Lab Directed Research and Development'' grant LDRD~10-043. BPS gratefully acknowledges a Goldhaber Distinguished Fellowship from Brookhaven Science Associates.

\appendix

\section{The IP-Sat model}\label{sec:ipsat}
 The impact parameter dependent dipole saturation model (IP-Sat)~\cite{Kowalski:2003hm} is a refinement of the Golec-Biernat--W\"usthoff dipole model~\cite{GolecBiernat:1998js,GolecBiernat:1999qd} to give the right perturbative limit when $\rt\rightarrow0$~\cite{Bartels:2002cj}. It is equivalent to the expression derived in the classical effective theory of the CGC, to leading logarithmic accuracy~\cite{McLerran:1998nk}. 

The proton dipole cross-section in this model is expressed as 
\begin{eqnarray}
\dsigmap(\rt,x,\bt)&=&2\left[1-\exp\left(-F(\rt, x, \bt)\right)\right]
\label{eq:ipsat-dipole}
\end{eqnarray}
with
\begin{eqnarray}
F(\rt,x,\bt)&=&\frac{\pi^{2}}{2N_{c}}\rt^{2}\alpha_{S}(\tilde{\mu}^{2}) xg(x,\tilde{\mu}^{2})T_p(\bt) .
\label{eq:ipsat-F}
\end{eqnarray}
Here the scale $\tilde{\mu}^{2}$ is related to dipole radius $\rt$ as
\begin{equation}
\tilde{\mu}^{2}=\frac{4}{\rt^{2}}+\tilde{\mu}_{0}^{2}\,,
\end{equation}
and the leading order expression for the running coupling is given by Eq.\,(\ref{eq:runningalpha}).
The model includes saturation as eikonalized power corrections to the DGLAP leading twist expression and may be valid in the regime where logs in $Q^2$ dominate logs in $x$. The saturation scale for a fixed impact parameter is determined self--consistently by requiring that the dipole amplitude (within brackets in eq.~\ref{eq:ipsat-dipole})  have the magnitude ${\cal N}(x,r_S,\bt)= 1-e^{-1/2}$, with $\qssp = 2/r_S^2$. We note that there is an overall logarithmic uncertainty in the determination of $\qssp (x,\bt)$.

For each value of the  dipole radius, the gluon density $xg(x,\tilde{\mu}^{2})$ is evolved from $\tilde{\mu}_{0}^{2}$ to $\tilde{\mu}^{2}$ using LO DGLAP evolution equation without quarks, 
\begin{equation}
\frac{\partial xg(x,\tilde{\mu}^{2})}{\partial \log \tilde{\mu}^{2}} = \frac{\alpha_{S}(\tilde{\mu}^{2})}{2\pi}\int \limits_{x}^{1}
dz P_{gg}(z)\frac{x}{z}g\left(\frac{x}{z},\tilde{\mu}^{2}\right) 
\end{equation}
Here the gluon splitting function with $N_{f}$ flavors and $C_{A}$=3 and $T_{R}$=1 is 
\begin{align}
 P_{gg}(z)=&6\left[\frac{z}{(1-z)+}+\frac{1-z}{z}+z(1-z)\right]\nonumber\\
&+\left(\frac{11}{2}-\frac{N_{f}}{3}\right)\delta(1-z)\,.
\end{align}
The initial gluon density at the scale $\tilde{\mu}^{2}_{0}$ is taken to be of the form
\begin{equation}
 xg(x,\tilde{\mu}^{2}_{0})=A_{g}x^{-\lambda_{g}}(1-x)^{5.6}\,.
\end{equation}
An important feature of the IP-Sat model is the b-dependence of the dipole cross-section, which is introduced through a gluon density profile function $T(b)$. 
This profile function is normalized to unity and is chosen to have the Gaussian form  
\begin{equation}
 T_{p}(\bt)=\frac{1}{2\pi B_{G}} \exp\left({-\bt^{2}\over 2B_{G}}\right)\,,
\label{eq:IPsat-imp-par}
\end{equation}
where $B_G$ is a parameter fit to the HERA diffractive data. This corresponds to $\langle b^2\rangle = 2 B_G$, the average squared {\it gluonic} radius of the proton, which corresponds to a very small transverse  radius $\sim 0.5$ fm and a three-dimensional radius of $\sim 0.61$ fm~\cite{Caldwell:2010zza}.

\vspace{-0.3cm}
\bibliography{spires}

\end{document}